\documentclass[aps,prd,twocolumn,showpacs]{revtex4}
\usepackage{epsfig,psfrag,amssymb}

\begin{document}

\title{On the exact gravitational lens equation in spherically symmetric
and static spacetimes}

\author{Volker Perlick}
\email[]{vper0433@itp.physik.tu-berlin.de}
\affiliation{Institut f{\"u}r Theoretische Physik. Universit{\"a}t zu K{\"o}ln\\
Z{\"u}lpicher Str. 77, 50937 K{\"o}ln, Germany }
\altaffiliation{Permanent address: Institut f{\"u}r Theoretische Physik, 
TU Berlin, Sekr. PN 7-1, 10623 Berlin, Germany}

\begin{abstract}
Lensing in a spherically symmetric and static spacetime is considered,
based on the lightlike geodesic equation without approximations. After
fixing two radius values $r_O$ and $r_S$, lensing for an observation
event somewhere at $r_O$ and static light sources distributed at $r_S$
is coded in a lens equation that is explicitly given in terms of 
integrals over the metric coefficients. The lens equation relates two
angle variables and can be easily plotted if the metric coefficients have
been specified; this allows to visualize in a convenient way
all relevant lensing properties, giving image positions, apparent 
brightnesses, image distortions, etc. Two examples are treated: 
Lensing by a Barriola-Vilenkin monopole and lensing by an Ellis 
wormhole.   
\end{abstract}

\maketitle

\section{Introduction\label{sec:intro}}
Theoretical work on gravitational lensing is traditionally done in a 
quasi-Newtonian approximation formalism, see,  e.g., Schneider, Ehlers 
and Falco \cite{schneider-ehlers-falco-92} or Petters, Levine and 
Wambsganss \cite{petters-levine-wambsganss-2001}, which is based, among
other things, on the approximative assumptions that the gravitational 
field is weak and that the bending angles are small. Under these assumptions, 
lensing is described in terms of a ``lens equation'' that determines a 
``lens map'' from a ``deflector plane'' to a ``source plane'', thereby 
relating image positions on the observer's sky to source positions. Although 
for all practical purposes up to now this formalism has proven to be very 
successful, there are two motivations for doing gravitational lens theory 
beyond the quasi-Newtonian approximation. First, from a methodological point 
of view it is desirable to investigate qualitative features of lensing, such 
as criteria for multiple imaging or for the formation of Einstein rings, 
in a formalism without approximations, as far as possible, to be sure that
these features are not just reflections of the approximations.  Second, 
lensing phenomena where strong gravitational fields and large bending angles 
are involved are no longer as far away from observability as they have been 
a few years ago. In particular, the discovery that there is a black hole at 
the center of our galaxy \cite{falcke-hehl-2003}, and 
probably at the center of most galaxies, has brought the matter of lensing in 
strong gravitational fields with large bending angles closer to practical 
astrophysical interest. If a light ray comes sufficiently close to a black 
hole, the bending angle is not small; in principle, it may even become 
arbitrarily large, corresponding to the light ray making arbitrarily many 
turns around the black hole. Unboundedly large bending angles also occur e.g. 
with wormholes; the latter are more exotic than black holes, in the sense 
that up to now there is no clear evidence for their existence, but 
nonetheless considered as hypothetical candidates for lensing by many authors.

If one wants to drop the assumptions of weak fields and small angles, gravitational
lensing has to be based on the lightlike geodesic equation in a general-relativistic
spacetime, without approximations. In this paper we will discuss this issue for the
special case of a spherically symmetric and static spacetime. In view of applications,
this includes spherical non-rotating stars and black holes, and also more exotic 
objects such as wormholes and monopoles with the desired symmetries. The main goal 
of this paper is to demonstrate that in this case lensing without approximations
can be studied, quite conveniently, in terms of a lens equation that is 
not less explicit than the lens equation of the quasi-Newtonian formalism.

Lensing without weak-field or small-angle approximations was pioneered by Darwin
\cite{darwin-58,darwin-61} and by Atkinson  \cite{atkinson-65}. Whereas Darwin's 
work is restricted to the Schwarz\-schild spacetime 
throughout, Atkinson derives all relevant formulas for an unspecified spherically 
symmetric and static spacetime before specializing to the Schwarzschild spacetime
in Schwarzschild and in isotropic coordinates. All important features of Schwarzschild
lensing are clearly explained in both papers. In particular, they discuss the occurrence 
of infinitely many images, corresponding to light rays making arbitrarily many turns
around the center and coming closer and closer to the light sphere at $r=3m$.
However, they do not derive anything like a lens equation.

The notion of a lens equation without weak-field or small-angle approximations 
was brought forward much later by Frittelli and Newman \cite{frittelli-newman-99}.
It is based on the idea of parametrizing the light cone of an arbitrary observation 
event in a particular way. For a general discussion of this idea and of the resulting 
``exact gravitational lens map'' in arbitrary spacetimes the reader may consult 
Ehlers, Frittelli and Newman \cite{ehlers-frittelli-newman-2003} or Perlick 
\cite{perlick-2001}. 
Here we are interested only in the special case of a spherically symmetric
and static spacetime. Then the geodesic equation is completely integrable
and the exact lens equation of Frittelli and Newman can be written 
quite explicitly. One can evaluate this equation from the spacetime 
perspective, as has been demonstrated by Frittelli, Kling and Newman
\cite{frittelli-kling-newman-2000} for the case of the Schwarzschild 
spacetime, thereby getting a good idea of the geometry of the light 
cone. Here we will use an alternative representation, using 
the symmetry for reducing the dimension of the problem.  After
fixing two radius values $r_O$ and $r_S$, lensing for an observation
event somewhere at $r_O$ and static light sources distributed at $r_S$
is coded in a lens equation, explicitly given in terms of integrals
over the metric coefficients, that relates two angles to each other. This 
representation results in a particularly convenient method of visualizing 
all relevant lensing properties, as will be demonstrated with two examples. 

The lens equation discussed in this paper should be compared with the lens 
equation for spherically symmetric and static spacetimes that was introduced 
by Virbhadra, Narasimha and Chitre \cite{virbhadra-narasimha-chitre-98} and 
then, in a modified form, by Virbhadra and Ellis \cite{virbhadra-ellis-2000}.
The Virbhadra-Ellis lens equation has found considerable interest. It was applied 
to the Schwarzschild spacetime \cite{virbhadra-ellis-2000} and later also to other 
spherically symmetric and static spacetimes, e.g. to a boson star by D\c{a}browski 
and Schunck \cite{dabrowski-schunck-2000}, to a fermion star by Bili{\'c}, Nikoli{\'c}
and Viollier \cite{bilic-nikolic-viollier-2000}, to spacetimes with naked singularities 
by Virbhadra and Ellis \cite{virbhadra-ellis-2002}, to the Reissner-Nordstr\"om 
spacetime by Eiroa, Romero and Torres \cite{eiroa-romero-torres-2002} and to a 
Gibbons-Maeda-Garfinkle-Horowitz-Strominger black hole by Bhadra \cite{bhadra-2003}. 
In the last two papers, the authors concentrate on light rays that make several
turns around the center and they use analytical methods developed by Bozza
\cite{bozza-2002}. The Virbhadra-Ellis lens equation takes an intermediary position 
between the exact lens equation and the quasi-Newtonian approximation. It makes no 
assumptions as to the smallness of bending angles, but it does make approximative 
assumptions as to the position of light sources and observer. For the Virbhadra-Ellis 
lens equation to be valid the spacetime must be asymptotically flat for $r \to 
\infty$ and both observer and light sources must be at positions where $r$ is 
large; moreover, one has to restrict to light sources close to the radial line 
opposite to the observer position, i.e., to the case that there is only a small 
misalignment. (The question of how one can free oneself from the latter assumption 
was addressed by D\c{a}browski and Schunck \cite{dabrowski-schunck-2000} and by 
Bozza \cite{bozza-2003}.) The lens equation to be discussed in the present 
paper is not restricted to the asymptotically flat case, and it makes no 
restriction on the position of light sources or observer. 


\section{Derivation of the lens equation\label{sec:def}}
We consider an arbitrary spherically symmetric and static spacetime.
For our purpose it will be advantageous to write the metric in the
form
\begin{equation}\label{eq:g}
  g \, = \, A(r)^2 \, \Big(  \, S(r)^2 \, dr^2 
  \, + \, R(r)^2 \, ( d \vartheta ^2 + {\mathrm{sin}} ^2 \vartheta 
  \, d \varphi ^2) \, - \, dt^2 \, \Big) \, .
\end{equation} 
Here $\varphi$ and $\vartheta$ are the standard coordinates on the
sphere, $t$ ranges over $\mathbb{R}$ and $r$ ranges over an open 
interval $\; ]\, r_{\mathrm{min}} \, , \, r_{\mathrm{max}} \, [ \;$ 
where $- \infty \le r_{\mathrm{min}} < r_{\mathrm{max}} \le \infty$. 
We assume that the functions $A$, $S$, and $R$ are strictly positive 
and (at least piecewise) differentiable on the interval 
$\; ]\, r_{\mathrm{min}} \, , \, r_{\mathrm{max}} \, [ \;$. 
As the lightlike geodesics are not affected
by the conformal factor $A(r)^2$ (apart from their parametrizations), the 
lens equation will depend on the metric coefficients $S(r)$ and $R(r)$ only.
We will see below that many qualitative features of
the lens equation are determined by the coefficient $R(r)$ alone.

\begin{figure}
    \psfrag{ppp}{\small $r=r_O$} 
    \psfrag{qqq}{\small $r=r_S$}  
    \psfrag{f}{\small $\varphi = 0$}  
    \psfrag{T}{\small $\Theta$} 
    \psfrag{g}{\small $\Phi$}  
    \psfrag{s}{\small $\psi$}  
    \psfrag{a}{\small $\partial _r$}  
    \psfrag{b}{\small $\partial _{\varphi}$}  
 \epsfig{file=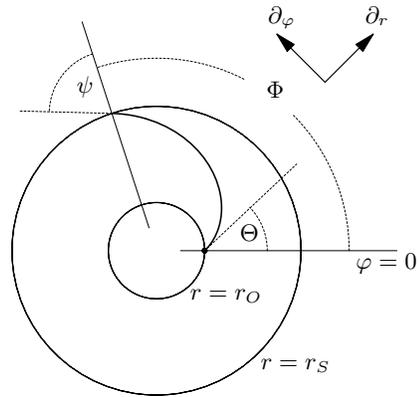,width=0.6\linewidth} 
 \caption{The figure shows the plane $\vartheta = \pi /2$. The observer, 
 indicated by a dot, is situated at $r = r_O$, $\varphi = 0$, the light sources
 are distributed at $r = r_S$. The lens equation relates the angle $\Theta$ which
 gives the image position on the observer's sky to the angle $\Phi$ which gives 
 the source position in the spacetime. The angle $\psi$ indicated in this picture 
 is irrelevant for the lens equation but will be needed in Subsection 
 \ref{subsec:ang}.\label{fig:lensmap}}
\end{figure}

For introducing our lens equation we have to fix two radius values $r_O$ and 
$r_S$ between $r_{\mathrm{min}}$ and $r_{\mathrm{max}}$.
The index $O$ stands for ``observer'', the index $S$ stands for 
``source''. We think of an observer at $r=r_O$, $\varphi = 0$, 
$\vartheta = \pi /2$. It is our goal to determine the appearance,
on the observer's sky, of static light sources distributed on the sphere 
$r=r_S$.   

By symmetry, we may restrict to the plane $\vartheta = \pi /2$. We consider 
past-oriented lightlike geodesics that start at time $t=0$ at the observer
and terminate, at some time $t=-T<0$ which depends on the geodesic, somewhere 
on the sphere $r = r_S$. To each of those light rays we assign the angle 
$\Theta$, measured at the observer between the ray's tangent and the direction 
of $\partial _r$, and the angle $\Phi$, swept out by the azimuth coordinate 
along the ray on its way from the observer to the source, see Figure 
\ref{fig:lensmap}. The desired lens equation is an equation of the form 
$\mathcal{F} (\Theta , \Phi )=0$ which relates image positions on the 
observer's sky, given by $\Theta$, to source positions in the spacetime, given 
by $\Phi$ modulo $2 \pi$. We restrict $\Theta$ to values between $- \pi$ and 
$\pi$; then $|\Theta |$ can be viewed as a colatitude coordinate on the observer's 
celestial sphere. By symmetry, $\mathcal{F} (\Theta, \Phi ) = 0$ must be equivalent
to $\mathcal{F} (- \Theta , - \Phi )=0$. For a given angle $\Theta$, neither existence 
nor uniqueness of an angle $\Phi$ with $\mathcal{F} (\Theta , \Phi )=0$ is guaranteed. 
Existence fails if the respective light ray never meets the sphere $r = r_S$; uniqueness 
fails if it meets this sphere several times. In the latter case the observer sees two 
or more images of light sources at $r_S$ at the same point on the sky, one behind the 
other. We will refer to images which are covered by other images as to ``hidden images''. 
The lens equation can be solved for $\Phi$, thereby giving a \emph{lens map} $\Theta 
\mapsto \Phi$, only if hidden images do not exist (or are willfully ignored).

To work out the lens equation we have to calculate the lightlike geodesics in 
the plane $\vartheta = \pi /2$ of the metric (\ref{eq:g}), which is an elementary
exercise. As a conformal factor has no influence on the lightlike geodesics (apart 
from their parametrization), they are solutions of the Euler-Lagrange equations 
of the Lagrangian $\mathcal{L} = \frac{1}{2} \big( S(r)^2 \dot{r} {}^2 + R(r)^2 
\dot{\varphi} {}^2 - \dot{t}{}^2 \big)$, i.e.
\begin{equation}\label{eq:ELr}
  \big( S(r)^2 \dot{r} \big) \dot{\,} \, = \, 
  S(r) \, S \, '(r) \, \dot{r} {}^2 
  \, + \, R(r) \, R \, '(r) \, \dot{\varphi} {}^2  \; ,
\end{equation}
\begin{equation}\label{eq:ELphi}   
  \big( R(r)^2 \dot{\varphi} \big) \dot{\,} \, = \, 0 \; ,
\end{equation}
\begin{equation}\label{eq:ELt}
  \ddot{t} \, = \, 0  \; ,
\end{equation}
where an overdot denotes differentiation with respect to the curve parameter $s$.
As an aside we mention that, by (\ref{eq:ELr}), a circular light ray exists at radius 
$r_p$ if and only if $R'(r_p)=0$. Comparing this condition with the equivalent but
less convenient eq. (33) in Atkinson's article \cite{atkinson-65} shows that it is
advantageous to write the metric in the form (\ref{eq:g}). 
The relevance of circular light rays in view of lensing was discussed by Hasse and 
Perlick \cite{hasse-perlick-2002}, also see Claudel, Virbhadra and Ellis 
\cite{claudel-virbhadra-ellis-2001} for related results. 

To get the past-oriented light ray that starts at time $t=0$ at the observer in
the direction determined by the angle $\Theta$ we have to impose the initial
conditions
\begin{equation}\label{eq:inr}
  r |_{s=0} \, = \, r_O \; , \qquad 
  \dot{r} |_{s=0} \, = \, \frac{\mathrm{cos} \, \Theta }{S(r_O)} \; ,
\end{equation}
\begin{equation}\label{eq:inphi}   
  \varphi |_{s=0} \, = \, 0 \; , \qquad 
  \dot{\varphi} |_{s=0} \, = \, \frac{\mathrm{sin} \, \Theta }{R(r_O)} \; ,
\end{equation}
\begin{equation}\label{eq:int}
  t |_{s=0} \, = \, 0 \; , \qquad 
  \dot{t} |_{s=0} \, = \, - \, 1 \; .
\end{equation}
For each $\Theta$, the initial value problem (\ref{eq:ELr}), (\ref{eq:ELphi}),
(\ref{eq:ELt}),(\ref{eq:inr}), (\ref{eq:inphi}), (\ref{eq:int}) has a unique 
maximal solution 
\begin{equation}\label{eq:fh}
  r= \mathsf{r} (\Theta , s) \; , \qquad \varphi = \phi (\Theta , s) \; , \qquad t = -s 
\end{equation}
where $s$ ranges from 0 up to some $s_{\mathrm{max}} (\Theta )$. 
Every image on the oberver's sky of a light source at $r_S$ corresponds to a pair
$(\Theta , \Phi )$ such that 
\begin{equation}\label{eq:el}
 r_S= \mathsf{r} (\Theta , T) \qquad \text{and} \qquad \Phi = \phi (\Theta , T)
\end{equation}
with some parameter value $T \in \; ] \, 0 , s_{\mathrm{max}} (\Theta )\, [ \:$. 
In other words, we get the desired lens equation $\mathcal{F} (\Theta , \Phi )=0$ 
if we eliminate $T$ from the two equations (\ref{eq:el}).
 
We get an explicit expression for the lens equation, and for the travel 
time $T$, by writing the functions $\mathsf{r} (\Theta , s)$ and 
$\phi (\Theta , s)$ in terms of integrals. From the constant of motion 
\begin{equation}\label{eq:null}
   S(r)^2 \, {\dot{r}}{}^2 \, + 
  \, R(r)^2 \, {\dot{\varphi}}{}^2 \, - \, {\dot{t}}{}^2 \,= \, 0   
\end{equation}
we find, with the help of (\ref{eq:ELphi}), (\ref{eq:ELt}),(\ref{eq:inphi}), 
(\ref{eq:int}),
\begin{equation}\label{eq:rs}
  S(r)^2 \, R(r)^2 \, \dot{r} {}^2 \, = \, 
  R(r)^2 - R(r_O)^2 {\mathrm{sin}} ^2 \Theta   \; .
\end{equation}  
If $\dot{r}$ does not change sign, integration of (\ref{eq:rs}) yields
\begin{equation}\label{eq:intrs}
  s \, = \, \frac{|\mathrm{cos} \, \Theta |}{\mathrm{cos} \, \Theta} \,
  \int_{r_O}^{ \mathsf{r} (\Theta,s)} \frac{R(r) \, S(r) \, dr}{\sqrt{R(r)^2-R(r_O)^2
  \, \mathrm{sin} ^2 \Theta}} \: .
\end{equation}  
With $\mathsf{r} (\Theta ,s)$ known, $\varphi = \phi (\Theta , s)$ is 
determined by integrating (\ref{eq:ELphi}) with (\ref{eq:inphi}),
\begin{equation}\label{eq:intphis}
  \phi (\Theta , s) \, = \,
  \int_{0}^{s} \frac{ R(r_O) \, \mathrm{sin} \, \Theta \,  d \tilde{s}}{
  R \big( \mathsf{r} (\Theta , \tilde{s} ) \big)^2} \: .
\end{equation}
(\ref{eq:intphis}) can be rewritten as an integral over $r$, with 
$\dot{r} = dr/ds$ substituted from (\ref{eq:rs}). This gives us the lens
equation in the form
\begin{equation}\label{eq:Phi}
  \Phi \, = \, \frac{|\mathrm{cos} \, \Theta |}{\mathrm{cos} \, \Theta} \,
  \int_{r_O}^{r_S} \frac{ R(r_O) \, \mathrm{sin} \, \Theta \,  S(r) \, d r}{
  R(r) \, \sqrt{R(r)^2-R(r_O)^2 \mathrm{sin}^2 \Theta}} \: .
\end{equation}
If $\dot{r}$ changes sign, (\ref{eq:intrs}) has to be
replaced by a piecewise integration. Similarly, the substitution from
the $s$-integration in (\ref{eq:intphis}) to an $r$-integration must
be done piecewise. In this case, the lens equation is not of the 
form (\ref{eq:Phi}); in particular, it is not guaranteed that the lens
equation can be solved for $\Phi$. In any case, we get exact integral 
expressions for the lens equation, and for the travel time $T$, from 
which all relevant lensing features can be determined in a way that is 
not less explicit than the quasi-Newtonian approximation formalism.
This will be demonstrated by two examples in Section \ref{sec:examples}.
In Subsection \ref{subsec:mono} we treat a particularly simple example
where the metric coefficients $R(r)$ and $S(r)$ are analytic and the
integral (\ref{eq:intrs}) can be explicitly calculated in terms
of elementary functions. In this case it suffices to calculate
(\ref{eq:intrs}) for arbitrarily small $s$ to get the whole
function $r(\Theta, s)$ by maximal analytic extension; i.e., in
this case it is not necessary to determine the points where $\dot{r}$
changes sign and to perform a piecewise integration.   


\section{Discussion of the lens equation\label{sec:dis}}
In the first part of this section we want to discuss for which values of 
$\Theta$ the lens equation $\mathcal{F} (\Theta , \Phi ) = 0$ admits a 
solution. In other words, we want to determine which part of the
observer's sky is covered by the light sources distributed at $r=r_S$.
We restrict to the case $r_O<r_S$. (The results for the case $r_O > r_S$ 
follow immediately from our discussion; we just have to make a coordinate 
transformation $r \to -r$ and, correspondingly, to change $\Theta$ into 
$\pi - \Theta$. The case $r_O = r_S$ can be treated by a limit procedure.)

For a light ray with one end-point at $r_O$ and the other at $r_S$ the
right-hand side of (\ref{eq:rs}) must be non-negative for all $r$ between
$r_O$ and $r_S$. This condition restricts the possible values of $\Theta$
by $\mathrm{sin} ^2 \Theta \le \mathrm{sin} ^2 \delta _I$ where 
\begin{equation}\label{eq:CI}
  \mathrm{sin} \, \delta _I \, =  \, {\mathrm{inf}} \, 
  \big\{ \, R(r)/R(r_O) \, \big| \, r_O < r < r_S \, \big\} \; .
\end{equation}
Note that our assumptions guarantee that this infimum is strictly positiv,
$0 < \delta _I \le \pi /2$. 

\begin{figure}
    \psfrag{aaaa}{\small $\Theta = 0 $} %
    \psfrag{bbbb}{\small $\Theta = \pi /2$} %
    \psfrag{cccc}{\small $\Theta = \pi \simeq - \pi$} %
    \psfrag{dddd}{\small $\Theta = -\pi /2$} %
    \psfrag{mm}{\small $\delta _I$} %
    \psfrag{kk}{\small $\delta _{II}$} %
 \epsfig{file=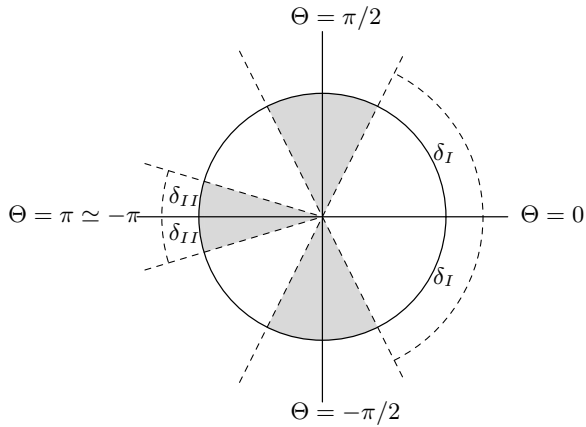,width=0.95\linewidth} %
 \caption{In the case $r_O < r_S$, the light sources cover the non-shaded region on
 the observer's sky, with $\delta _I$ and $\delta _{II}$ given by (\ref{eq:CI}) and 
 (\ref{eq:CII}), respectively. (For $r_S < r_O$ one gets an analogous picture, with 
 $\Theta =0$ and $\Theta = \pi \simeq - \pi$ interchanging their roles.) If the 
 spacetime is asymptotically flat and if $r_S$ is sufficiently large, this picture 
 gives the socalled ``escape cones'' which have been calculated by Synge 
 \cite{synge-66} for the Schwarzschild spacetime and by Pande and Durgapal 
 \cite{pande-durgapal-86} for an unspecified asymptotically flat spherically 
 symmetric and static spacetime.\label{fig:sky}}
\end{figure}

Furthermore, a light ray with $\pi /2 < | \Theta | < \pi$ can arrive at $r_S$
only if it passes through a minimal radius value $\rho (\Theta ) < r_O$.
As (\ref{eq:rs}) requires $R \big( \rho ( \Theta ) \big) ^2 = R(r_O)^2
\mathrm{sin} ^2 \Theta$, this can be true only if $\mathrm{sin} ^2 \Theta \ge
\mathrm{sin} ^2 \delta _{II}$ where 
\begin{equation}\label{eq:CII}
  \mathrm{sin} \, \delta _{II} \, = \, {\mathrm{inf}} \, 
  \big\{ \, R(r) / R(r_O) \, \big| \, r_{\mathrm{min}} < r < r_O \, \big\} \; ,
\end{equation}
$0 \le \delta _{II} \le \pi /2$.
So in general the light sources at $r_S$ cover on the observer's sky a 
disk of angular radius $\delta _I$ around the pole $\Theta = 0$ and, if $\delta _{II}
< \delta _I$, in addition a ring of angular width $\delta _I - \delta _{II}$ around 
the pole $\Theta = \pi \simeq - \pi$, see Figure \ref{fig:sky}. The two domains
join if $\delta _I = \pi /2$.

We see that the allowed values of $\Theta$ are determined by the metric
coefficient $R$ alone. We will now demonstrate that $R$ alone also determines
the occurrence or non-occurrence of hidden images. Hidden images occur 
if a light ray from $r_O$ intersects the sphere $r=r_S$ at least two times; 
between these two intersections it must pass through a
maximal radius $\sigma (\Theta ) > r_S$ which, by (\ref{eq:rs}), has 
to satisfy $R \big( \sigma ( \Theta ) \big) ^2 = R(r_O)^2 \mathrm{sin} ^2 \Theta$.
Such a radius $\sigma (\Theta )$ exists for all $\Theta$ with 
$\mathrm{sin} ^2 \Theta > C^2$ where
\begin{equation}\label{eq:C}
  C  \, = \, {\mathrm{inf}} \, 
  \big\{ \, R(r) / R(r_O) \, \big| \, r_S  < r < r_{\mathrm{max}}\, \big\} \; .
\end{equation}
As $\Theta$ is restricted by $\mathrm{sin} ^2 \Theta \le \mathrm{sin} ^2 \delta _I$, 
hidden images cannot occur if $C \ge \mathrm{sin} \, \delta _I$. The latter 
condition is satisfied in asymptotically flat spacetimes, where $R(r) \to \infty$ 
for $r \to r_{\mathrm{max}} \, (= \infty )$, if we choose $r_S$ sufficiently large. 
This is the reason why in the more special situation of the Virbhadra-Ellis lens 
equation \cite{virbhadra-ellis-2000} hidden images cannot occur.

In the rest of this section we discuss the question of multiple imaging and
the occurrence of Einstein rings. For a light source at $r=r_S$, $\varphi = \varphi _0$, 
$\vartheta = \pi /2$ with $0 < | \varphi _0 | < \pi$, images on the observer's sky 
are in one-to-one correspondence with solutions $\Theta$ of the equation
\begin{equation}\label{eq:mult}
  \mathcal{F} \big( \Theta , \varphi _0 \, + \, 2 \, n \, \pi \big)
  \, = \, 0 
\end{equation}
with $n \in \mathbb{Z}$. We call the integer $n$ 
the ``winding number'' of the corresponding light ray. An image with
$n = 0$ is called ``primary'' and an image with $n=
- \varphi _0 / | \varphi _0 |$ is called ``secondary''. Images with other values of $n$
correspond to light rays that make at least one full turn and have been 
termed ``relativistic'' by Virbhadra and Ellis \cite{virbhadra-ellis-2000}. 
Note that different images of a light source may have the same 
winding number. 

If we send $\varphi _0$ to 0 or to $\pi \simeq - \pi$, solutions $\Theta$ of equation 
(\ref{eq:mult}) with $\mathrm{sin} \, \Theta \neq 0$ come in pairs $(\Theta, - \Theta )$. By 
spherical symmetry, every such pair gives rise to an Einstein ring. There are 
as many Einstein rings as the equation
\begin{equation}\label{eq:ring}
  \mathcal{F} ( \Theta ,  i \, \pi ) \, = \, 0 
\end{equation}
admits solutions with positive integers $i$. Even integers $i$ correspond
to Einstein rings of the source at $\varphi _0 = 0$, and odd integers
$i$ correspond to Einstein rings of the source at  $\varphi _0 = \pi \simeq
- \pi$. 


\section{Observables\label{sec:obs}}
To each solution $(\Theta , \Phi )$ of the lens equation we can assign 
redshift, travel time, apparent brightness and image distortion. 

\subsection{Redshift\label{subsec:red}}

The general redshift formula for static metrics (see, e.g., Straumann
\cite{straumann-84}, p. 97) specified to metrics of the form 
(\ref{eq:g}) says that the redshift $z$ is given by 
\begin{equation}\label{eq:z}
  1 \, + \, z \, = \, \frac{\, A(r_O) \,}{A(r_S)}
\end{equation}
if the observer's worldline is a $t$-line at $r=r_O$ and the source's 
worldline is a $t$-line at $r = r_S$. In our situation $r_O$ and $r_S$ are 
fixed, so the redshift is a constant. 


\subsection{Travel time\label{subsec:T}}
Recall that $(\Theta , \Phi )$ is a solution of the lens equation if and only if 
there is a parameter $T$ such that the equations (\ref{eq:el}) hold. This assigns
a travel time $T$ to each solution $(\Theta , \Phi )$ of the lens equation.
If there are no hidden images, the equation $\mathsf{r} (\Theta , T)=r_S$ gives 
$T$ as a single-valued function of $\Theta$.


\subsection{Angular diameter distance\label{subsec:ang}}
Quite generally, determination of the angular diameter distance requires solving the 
Sachs equations for the optical scalars along lightlike geodesics, see e.g. Schneider, 
Ehlers and Falco \cite{schneider-ehlers-falco-92}. For the Schwarzschild metric, this 
has been explicitly worked out by Dwivedi and Kantowski \cite{dwivedi-kantowski-72}. 
Their method easily carries over to arbitrary spherically symmetric and static spacetimes 
as was demonstrated by Dyer \cite{dyer-77}.  In what follows we give a reformulation of 
these results in terms of our lens equation. 

To that end we fix a solution $(\Theta, \Phi )$ of the lens equation and thereby a 
(past-oriented) light ray from the observer at $r_O$ to a light source at $r_S$. Around 
this ray, we consider an infinitesimally thin bundle of neighboring rays, with vertex at the 
observer. The angular diameter distance is defined as the square-root of the ratio between 
the cross-sectional area of this bundle at the light source and the opening solid angle 
at the observer. Owing to the symmetry of our situation there are two preferred 
spatial directions perpendicular to the ray: a radial direction (along a meridian 
on the observer's sky) and a tangential direction (along a circle of equal latitude on 
the observer's sky). Therefore, the angular diameter distance naturally comes about 
as a product of a radial part and a tangential part. 

To calculate the radial part, we consider the infinitesimally neighboring ray which 
corresponds to an infinitesimally neighboring solution $(\Theta + d \Theta , \Phi + d \Phi )$ 
of the lens equation, i.e. $d\Phi$ and $d \Theta$ satisfy
\begin{equation}\label{eq:dPhidTheta}
  \frac{\partial \mathcal{F}}{ \partial \Theta } (\Theta , \Phi ) \, + \,
  \frac{\partial \mathcal{F}}{ \partial \Phi } (\Theta , \Phi ) \,
  \frac{d \Phi }{d \Theta} \: = \: 0
\end{equation}
We define the \emph{radial angular diameter distance} as  
\begin{equation}\label{eq:Dr}
  D^r _{\mathrm{ang}} \, = \, A(r_S)\, R(r_S) \, 
  \mathrm{cos} \, \psi \, \frac{d \Phi}{d \Theta} 
\end{equation}
with $\psi$ given by Figure \ref{fig:lensmap}, i.e., $A(r_S)\, R(r_S) \, 
\mathrm{cos} \, \psi \, d \Phi$ measures, in the direction perpendicular 
to the original ray, how far the neighboring ray is away. By 
(\ref{eq:ELphi}), (\ref{eq:inphi}) and (\ref{eq:rs}), $\psi$ must satisfy
\begin{equation}\label{eq:psi}
  R(r_O) \, \mathrm{sin} \, \Theta \, = \, 
  R(r_S) \, \mathrm{sin} \, \psi \, .
\end{equation}   
With $\psi$ given by (\ref{eq:psi}) and $d\Phi / d\Theta$ given by (\ref{eq:dPhidTheta}), 
$D^r _{\mathrm{ang}}$ is determined by (\ref{eq:Dr}) for every solution $(\Theta, \Phi )$ 
of the lens equation. Note that $D^r _{\mathrm{ang}}$ is singular at those solutions
of the lens equation where $\partial \mathcal{F}/ \partial \Phi $ has a zero. If the 
lens equation can be solved for $\Phi$, we may view $D^r _{\mathrm{ang}}$ as a 
(single-valued) function of $\Theta$. 

To calculate the tangential part we consider an infinitesimally neighboring 
light  ray that results by applying a rotation around the axis $\varphi = 0$, 
$\vartheta = \pi /2$. Such rotations are generated by the Killing vector field 
$K= \mathrm{sin} \, \varphi \, \partial _{\vartheta} + \mathrm{cot} \, 
\vartheta \, \mathrm{cos} \, \varphi \, \partial _{\varphi}$. At points 
with $\vartheta = \pi /2$, this Killing vector field takes the form $K \, = 
\mathrm{sin} \, \varphi \, A(r) \, R(r) \, g( \partial _{\vartheta} , 
\partial _{\vartheta}) ^{-1/2} \, \partial _{\vartheta}$. Hence, if we rotate 
by an infinitesimal angle $d \beta$, the neighboring ray intersects the sphere 
$r=r_S$ at a distance $A(r_S) \, R(r_S) \, \mathrm{sin} \, \Phi  \, d \beta$ 
from the original ray. Relating this distance to the angle $\mathrm{sin} \, 
\Theta \, d \beta$ between the two rays at the observer gives the  
\emph{tangential angular diameter distance}   
\begin{equation}\label{eq:Dt}
  D^t _{\mathrm{ang}} \, = \, A(r_S) \, R(r_S) \, 
  \frac{\, \mathrm{sin} \, \Phi \,}{\mathrm{sin} \, \Theta} \; .
\end{equation}
By this equation, $D^t _{\mathrm{ang}}$ is uniquely determined for each solution 
$(\Theta , \Phi )$ of the lens equation. Again, $D^t _{\mathrm{ang}}$ may be viewed
as a function of $\Theta$ if the lens equation can be solved for $\Phi$.

$D^r _{\mathrm{ang}}$ and  $D^t _{\mathrm{ang}}$
together give the (averaged) \emph{angular diameter distance}
or \emph{area distance}
\begin{equation}\label{eq:D}
  D_{\mathrm{ang}} \, = \, \sqrt{\, |D^r _{\mathrm{ang}}
  \, D^t _{\mathrm{ang}} |\,} \; .
\end{equation}
Note that both $D^r _{\mathrm{ang}}$ and  $D^t _{\mathrm{ang}}$ may be negative. 
Images with $D^r _{\mathrm{ang}} \, D^t _{\mathrm{ang}} \, > \,0$ are said to have 
\emph{even parity} and images with $D^r _{\mathrm{ang}} \,D^t _{\mathrm{ang}} \, 
< \, 0$ are said to have \emph{odd parity}. Images with odd parity show the 
neighborhood of the light source side-inverted in comparison to images with even 
parity. 

A solution $(\Theta , \Phi )$ of the lens equation is called a \emph{radial 
critical point} if $D^r _{\mathrm{ang}} =0$ and a \emph{tangential 
critical point} if $D^t _{\mathrm{ang}}=0$. The latter condition is equivalent 
to $\mathrm{sin}\, \Phi =0$ and $\mathrm{sin} \, \Theta \neq 0$, i.e., to the 
occurrence of an Einstein ring. Note that (radial and tangential) critical 
points come in pairs, $(\Theta, \Phi )$ and $(- \Theta , - \Phi )$. Every 
such pair corresponds to a circle of equal latitude on the observer's sky 
which may be called a (radial or tangential) \emph{critical circle}, as in 
the quasi-Newtonian approximation formalism, see Schneider, Ehlers and
Falco \cite{schneider-ehlers-falco-92}, p. 233. In the quasi-Newtonian 
formalism one usually introduces the inverse magnification factors $1/ \mu ^r$ 
and $1/ \mu ^t$ as substitutes for $D^r _{\mathrm{ang}}$ and  $D^t _{\mathrm{ang}}\,$. 
In our situation, where there is no flat background metric, not even asymptotically, 
the magnification factors cannot be defined in a reasonable way, but working with 
$D^r _{\mathrm{ang}}$ and  $D^t _{\mathrm{ang}}\,$ is completely satisfactory.


\subsection{Luminosity distance\label{subsec:lum}}
In arbitrary spacetimes, the angular diameter distance $D_{\mathrm{ang}}$
is related to the (uncorrected) \emph{luminosity distance} $D_{\mathrm{lum}}$ 
by the universal formula $D_{\mathrm{lum}} = (1+z)^2 D_{\mathrm{ang}}$, see, e.g.,
Schneider, Ehlers and Falco \cite{schneider-ehlers-falco-92} , eq. (3.80).
With the redshift $z$ given by (\ref{eq:z}) and the angular diameter distance
$D_{\mathrm{ang}}$ given by (\ref{eq:D}), we find
\begin{equation}\label{eq:Dlum}
  D_{\mathrm{lum}} \, = \, \frac{\, A(r_O)^2 \,}{A(r_S)^2} \, \sqrt{\, |D^r _{\mathrm{ang}}
  \, D^t _{\mathrm{ang}} |\,} \; . 
\end{equation}
For an isotropically radiating light source with bolometric luminosity $L$,
the total flux at the observer is $L / (4 \pi D_{\mathrm{lum}}^2)$, see  again
Schneider, Ehlers and Falco \cite{schneider-ehlers-falco-92}, eq. (3.79).
Hence, if we distribute standard candles at $r=r_S$, their apparent 
brightness on the observer's sky is proportional to $D_{\mathrm{lum}}^{-2}$. 


\subsection{Image distortion\label{subsec:dist}}
$D^r _{\mathrm{ang}} $ and $D^t _{\mathrm{ang}}$ immediately
give the apparent distortion of images. For the sake of illustration, we may think of
small spheres, with infinitesimal diameter $d D$, distributed with their centers
at $r=r_S$. By definition of $D^r _{\mathrm{ang}}$ and $D^t _{\mathrm{ang}}$, each 
solution $( \Theta , \Phi )$ of the lens equation corresponds to an elliptical image
of such a sphere on the observer's sky, with the radial (meridional) diameter of 
the ellipse equal to $| d \Theta | = |d D / D^r _{\mathrm{ang}}|$ and with the tangential
(latitudinal) diameter of the ellipse equal to $| d \beta | = | d D / D^t _{\mathrm{ang}} |$. 
Thus, we may use the \emph{ellipticity}
\begin{equation}\label{eq:epsilon}
  \varepsilon \; = \; 
  \frac{\, |d \Theta | \, - \, |d \beta| \,}{
  | d \Theta| \, + \, |d \beta |} \; = \;
  \frac{\, | D^r _{\mathrm{ang}}| \, - \, | D^t _{\mathrm{ang}} | \, }{
  \,| D^r _{\mathrm{ang}}| \, + \, | D^t _{\mathrm{ang}} | }
\end{equation}
as a measure for image distortion. 


\section{Examples\label{sec:examples}}
\subsection{Lensing by a Barriola-Vilenkin monopole\label{subsec:mono}}
We consider the metric 
\begin{equation}\label{eq:monog}
    g \, = \, dr^2 \, + \, k^2 \, r^2 \, ( d \vartheta ^2 + {\mathrm{sin}} ^2 \vartheta 
  \, d \varphi ^2) \, - \, dt^2 \, 
\end{equation}
where $k$ is a positive constant. For $k=1$, this is just Minkowski spacetime in 
spherical coordinates. For $k < 1$, there is a deficit solid angle and a singularity 
at $r=0 \,$; the plane $t={\mathrm{const.}}$, $\vartheta = \pi /2$ has the geometry
of a cone. Similarly, for $k>1$ there is a surplus solid angle and a 
singularity at $r=0$. For $k \neq 1$, the metric is non-flat. The Einstein tensor 
has non-vanishing components $G_{tt} \, = \, - \, G_{rr} \, = \,(1-k^2)/r^2$, so 
the weak energy condition is satisfied (without a cosmological constant) if and 
only if $k \le 1$. In that case it was shown by Barriola and Vilenkin 
\cite{barriola-vilenkin-89} that the metric may be viewed as a model for the
spacetime around a monopole resulting from breaking a global $O(3)$ symmetry. 
To within the weak-field approximation, basic features of lensing by such a 
monopole were discussed in the original paper by Barriola and Vilenkin 
\cite{barriola-vilenkin-89} and also by Durrer \cite{durrer-94}. In what 
follows we give a detailed account in terms of our exact lens equation. Note that 
the Virbhadra-Ellis lens equation \cite{virbhadra-ellis-2000} is not applicable to 
this case, at least not without modification, because for $k \neq 1$ the spacetime 
is not asymptotically flat in the usual sense.

\begin{figure}
    \psfrag{aaa}{\small $- \pi$} %
    \psfrag{bbbbb}{\small $- \pi /2$} %
    \psfrag{ccc}{\small $\pi /2$} %
    \psfrag{d}{\small $\pi$} %
    \psfrag{UUUUU}{\small $- \pi / k$} %
    \psfrag{AAAA}{\small $-3 \pi$} %
    \psfrag{BBBB}{\small $-2 \pi$} %
    \psfrag{CCC}{\small $- \pi$} %
    \psfrag{DD}{\small $\pi$} %
    \psfrag{EEE}{\small $2 \pi$} %
    \psfrag{FFF}{\small $3 \pi$} %
     \psfrag{VVVV}{\small $\pi / k$} %
   \psfrag{X}{\small $\Theta$} %
    \psfrag{Y}{\small $\Phi$} %
 \epsfig{file=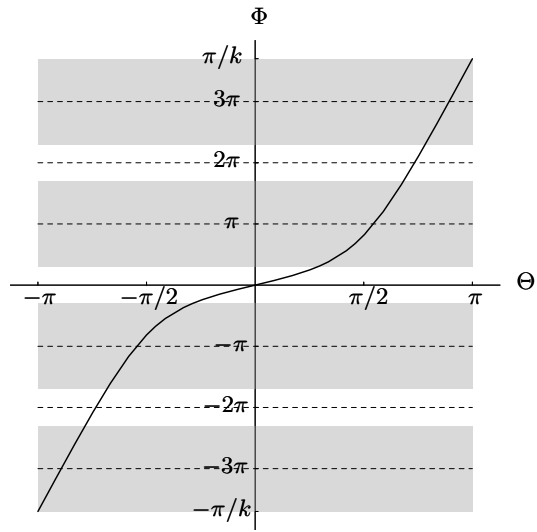,width=0.8\linewidth} %
 \caption{The lens equation (\ref{eq:monolens}) for the Barriola-Vilenkin 
 monopole with $k = 1/3.7$ and $r_O = 0.77 \, r_S$. Intersections with the dashed 
 lines indicate Einstein rings. Shading distinguishes sources with $N(k) + 1 = 4$
 images from sources with $N(k) = 3$ images. \label{fig:mono1}}
\end{figure}
 
\begin{figure}
    \psfrag{aaa}{\small $- \pi$} %
    \psfrag{sss}{\small $- \alpha$} %
    \psfrag{tt}{\small $\alpha$} %
    \psfrag{d}{\small $\pi$} %
    \psfrag{UUUUU}{\small $- \pi /k$} %
    \psfrag{AAAA}{\small $-3 \pi$} %
    \psfrag{BBBB}{\small $-2 \pi$} %
    \psfrag{CCC}{\small $-\pi$} %
    \psfrag{DD}{\small $\pi$} %
    \psfrag{EEE}{\small $2 \pi$} %
    \psfrag{FFF}{\small $3 \pi$} %
    \psfrag{VVVV}{\small $\pi /k$} %
    \psfrag{X}{\small $\Theta$} %
    \psfrag{Y}{\small $\Phi$} %
 \epsfig{file=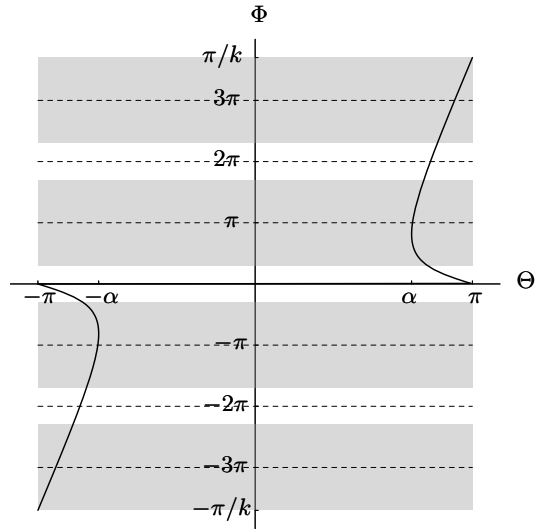,width=0.8\linewidth} 
 \caption{The lens equation (\ref{eq:monolens}) for the 
 Barriola-Vilenkin monopole with $k = 1/3.7$ and $r_S = 0.77 \, r_O$. 
 Other than in the case $r_O < r_S$, the lens equation cannot be solved for 
 $\Phi$, i.e., there are hidden images. The angle $\alpha$ is determined
 by $r_S \, = \, r_O \, \mathrm{sin} \, \alpha$ and $\pi /2 < \alpha < \pi$.
 The picture shows that all Einstein rings, indicated by intersections with 
 the dashed lines, are hidden. Shading distinguishes sources with 
 $N(k) + 1 = 4$ images from sources with $N(k) = 3$ images; only one 
 image of each source is non-hidden. \label{fig:mono2}}
\end{figure}

\begin{figure}
    \psfrag{aaa}{\small $- \pi$} %
    \psfrag{bbbbb}{\small $- \pi /2$} %
    \psfrag{ccc}{\small $\pi /2$} %
    \psfrag{d}{\small $\pi$} %
    \psfrag{PPPPPP}{\small $r_S + r_O \:$} %
    \psfrag{QQQQQQ}{\raisebox{-0.7ex}{\small $r_S - r_O \:$}} %
    \psfrag{X}{\small $\Theta$} %
    \psfrag{Y}{\small $D_{\mathrm{ang}}^r , D_{\mathrm{ang}}^t$} %
 \epsfig{file=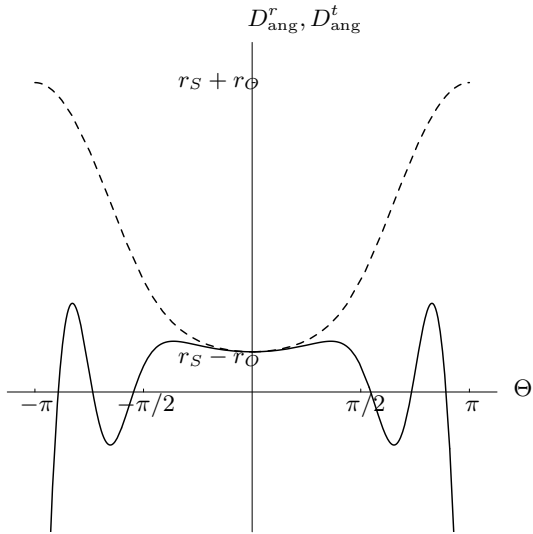,width=0.8\linewidth} 
 \caption{Tangential angular diameter distance $D_{\mathrm{ang}}^r $ (dashed) and 
 radial angular diameter distance $D_{\mathrm{ang}}^t $ (solid) as functions of 
 $\Theta$ for the Barriola-Vilenkin monopole with $k = 1/3.7$ and $r_O = 0.77 \, r_S$. 
 The dashed curve also gives the travel time $T = D_{\mathrm{ang}}^r$. Each 
 zero of $D_{\mathrm{ang}}^t$ indicates an Einstein ring. Where 
 $D_{\mathrm{ang}}^t$ and $D_{\mathrm{ang}}^r$ have the same sign, the 
 images have even parity; where $D_{\mathrm{ang}}^t$ and $D_{\mathrm{ang}}^r$ 
 have different signs, the images have odd parity, i.e., the neighborhood of 
 each light source is shown side-inverted.\label{fig:mono3}}
\end{figure}

Comparison of (\ref{eq:monog}) with (\ref{eq:g}) shows that the 
metric coefficients are given by
\begin{equation}\label{eq:monoR}
  A(r) \, = \, 1 \; , \qquad S(r) \, = \, 1 \; , \qquad R(r) \, = \, k \, r 
\end{equation}
on the interval $ r_{\mathrm{min}} = 0 < r < r_{\mathrm{max}} = \infty$.
With these metric coefficients, the integrals (\ref{eq:intrs}) and 
(\ref{eq:intphis}) can be calculated in an elementary fashion, yielding 
the solution to the initial value problem in the form
\begin{equation}\label{eq:monors}
  \mathsf{r} (\Theta , s) \, = \, 
  \sqrt{\, r_O^2\, + \,2 \, r_O \, s \, \mathrm{cos} \, \Theta \, + \, s^2 \,}
  \: ,
\end{equation}
\begin{equation}\label{eq:monophis}
  \phi (\Theta , s) \, = \, \frac{1}{k} \,
  \frac{\, \mathrm{sin} \, \Theta \, }{| \mathrm{sin} \, \Theta |} \,
  \mathrm{arccos} \Big(\frac{\, r_O \, + \, s \, \mathrm{cos} \, \Theta}{
  \sqrt{\, r_O^2 \, + \, 2 \, r_O \, s \, \mathrm{cos} \, \Theta \, + \, s^2 \,}} \Big)
  \: .
\end{equation}
For $\mathrm{sin} \, \Theta \, \neq \, 0 \,$, $s$ ranges from 0 to $\infty$, so 
$| \varphi | = | \phi (\Theta , s)|$ ranges from 0 to $|\Theta | /k$. 
Eliminating $T$ from the two equations (\ref{eq:el}) gives the lens equation,
\begin{equation}\label{eq:monolens}
  r_S \, \mathrm{sin} ( \Theta - k \Phi )\, - \, 
  r_O \, \mathrm{sin} \, \Theta \: = \: 0 \; ,
\end{equation}
which is to be considered on the domain
\begin{equation}\label{eq:monodom}
  - \pi < \Theta < \pi \: , \qquad k  | \Phi | < | \Theta | \: .  
\end{equation}
We restrict to the case that the integer $N(k)$ defined by $N(k) \le 1/k 
< N(k)+1$ is odd. The lens equation is plotted for the case $r_O < r_S$ in 
Figure \ref{fig:mono1} and for the case $r_O > r_S$ in Figure \ref{fig:mono2}. 
(For producing the pictures we have chosen $k$ such that $N(k)=3$.) In either 
case we find that there are $N(k)$ Einstein rings. For a light source at $r=r_S, 
\varphi = \varphi _0, \vartheta = \pi /2$ with $0 < |\varphi _0 | < \pi$ 
there are $N(k)$ images if $|\varphi _0 | \le  \big( 1 + N(k) - 1/k \big) 
\pi$ (non-shaded regions in Figures \ref{fig:mono1} and \ref{fig:mono2}) 
and  $N(k) + 1$ images otherwise (shaded regions in Figures \ref{fig:mono1} 
and \ref{fig:mono2}).

\begin{figure}
    \psfrag{aaa}{\small $- \pi$} %
    \psfrag{bbbbb}{\small $- \pi /2$} %
    \psfrag{ccc}{\small $\pi /2$} %
    \psfrag{d}{\small $\pi$} %
    \psfrag{BB}{\small $- 2$} %
    \psfrag{DD}{\small $\; 2$} %
    \psfrag{EE}{\small $\; 4$} %
    \psfrag{X}{\small $\Theta$} %
    \psfrag{Y}{\small $m= 2.5 \, \mathrm{log} _{10} ( D_{\mathrm{lum}} ^2 )+m_0$} %
 \epsfig{file=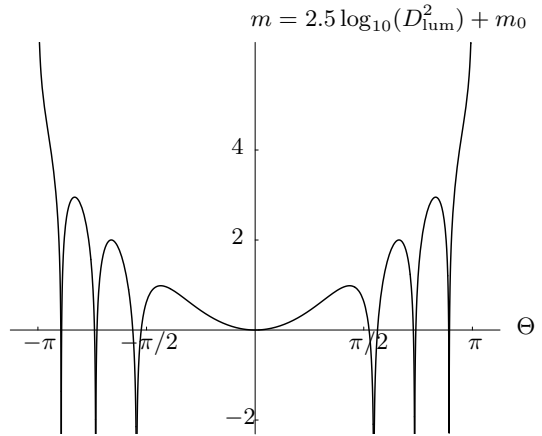,width=0.8\linewidth} 
 \caption{Apparent brightness of standard candles for the 
 Barriola-Vilenkin monopole with $k = 1/3.7$ and $r_O = 0.77 \, r_S$. 
 Instead of $D_{\mathrm{lum}}^{-2}$, which is proportional to the energy
 flux from point sources, we have plotted the magnitude $m= 2.5 \, \mathrm{log} _{10} 
 ( D_{\mathrm{lum}} ^2 ) + m_0$ used by astronomners. The constant $m_0$
 has been chosen such that $m=0$ at $\Theta =0$. Einstein rings are infinitely 
 bright in the ray optical treatment, $m= - \infty$.
 \label{fig:mono4}}
\end{figure}

In the following we concentrate on the case $r_O < r_S$. Then the lens equation can be 
solved for $\Phi$, giving a lens map
\begin{equation}\label{eq:monoPhi}
  \Theta \: \longmapsto \: \Phi  \, = \, 
  \frac{1}{k} \, \left( \, \Theta \, - \, \mathrm{arcsin}
  \Big( \, \frac{r_O}{r_S} \, \mathrm{sin} \, \Theta \, \Big) \right) 
\end{equation}
on the domain $- \pi < \Theta < \pi$, i.e., in the notation of Figure \ref{fig:sky}
we have $\delta _I = \pi /2$ and $\delta _{II} = 0$.  In the case $k=1$ (flat
spacetime), the lens map can be continuously extended into the  point $\Theta = \pi
\simeq - \pi$ (and by periodicity onto the full circle, 
$\Theta \in \mathbb{R}$ mod $2 \pi$). For $k \neq 1$ the ray with $\Theta = \pi
\simeq - \pi$ cannot pass through the singularity at $r=0$, so the lens map is
not to be extended beyond the open interval $\, \:]-\pi,\pi \, [ \: \,$. On this interval, 
$\Phi$ increases monotonously from $- \pi /k$ to  $\pi /k$, see Figure
\ref{fig:mono1}. Thus, the range of the lens map is independent of $r_O$ and $r_S$.
The ratio $r_O / r_S$ influences the shape of the graph of the lens map in the following
way. For  $r_O / r_S \to 0$ it becomes a straight line, $\Phi = \Theta /k$.
For $r_O / r_S \to 1$ it becomes a broken straight line, $\Phi = 0$ for 
$|\Theta| \le \pi /2$ and $\Phi = (2 \Theta - \pi )/k$ for $|\Theta - \pi | 
< \pi /2$. Note that linearity of the lens map implies that the angular distance 
on the observer's sky between consecutive images is the same for all light
sources. 

\begin{figure}
    \psfrag{aaa}{\small $- \pi$} %
    \psfrag{bbbbb}{\small $- \pi /2$} %
    \psfrag{ccc}{\small $\pi /2$} %
    \psfrag{d}{\small $\pi$} %
    \psfrag{X}{\small $\Theta$} %
    \psfrag{Y}{\small $\varepsilon$} %
 \epsfig{file=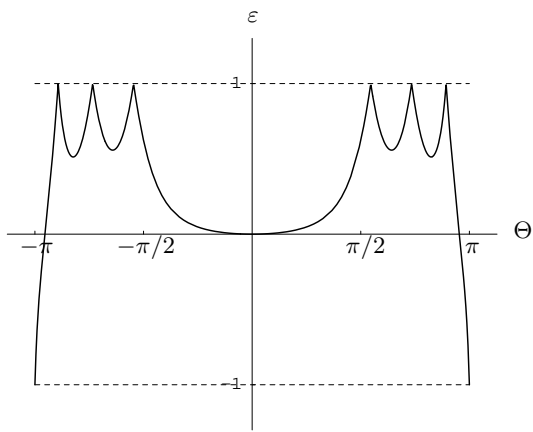,width=0.8\linewidth} 
 \caption{The ellipticity $\varepsilon$ for the Barriola-Vilenkin monopole 
 with $k = 1/3.7$ and $r_O = 0.77 \, r_S$. If small spheres are distributed with
 their centers at $r=r_S$, this function describes their distortion
 into ellipses on the observer's sky. $\varepsilon =0$ indicates 
 circular images, $\varepsilon =-1$ images with no tangential extension
 (radial critical points), and $\varepsilon =1$ images with no 
 radial extension (tangential critical points, i.e. Einstein
 rings). \label{fig:mono5}}
\end{figure}

As $\Phi$ is a (single-valued) function of $\Theta$, so are all observables. 
By evaluating the formulas derived in Section \ref{sec:obs} for the case at 
hand we find
\begin{equation}\label{eq:monoz}
  z \, = \, 0 \: ,
\end{equation}
\begin{equation}\label{eq:monoT}
  T \, = \, D_{\mathrm{ang}} ^r \, = \, r_O \, \mathrm{cos} \, \Theta \,
  + \, \sqrt{\, r_S^2 \, - \, r_O^2 \mathrm{sin} ^2 \Theta \,} \: ,
\end{equation}
\begin{equation}\label{eq:monoD}
  D_{\mathrm{ang}} ^t \, = \, \frac{k \, r_S}{\, \mathrm{sin} \, \Theta \,}
  \: \mathrm{sin} \left( \, \frac{1}{k} \, \Big( \, \Theta \, - \,
  \mathrm{arcsin} \big( \, \frac{r_O \, \mathrm{sin} \, \Theta \,}{r_S} \, \big) \Big) 
  \right) \: ,
\end{equation}
which gives $D _{\mathrm{lum}}$ and $\varepsilon$ as functions of $\Theta$
via (\ref{eq:Dlum}) and (\ref{eq:epsilon}). The observables are plotted in
Figures \ref{fig:mono3}, \ref{fig:mono4} and \ref{fig:mono5}.


\subsection{Lensing by an Ellis wormhole\label{subsec:worm}}
The metric
\begin{equation}\label{eq:wormg}
  g \, = \, dr^2 \, + \, (r^2+a^2) \, 
  ( d \vartheta ^2 + {\mathrm{sin}} ^2 \vartheta 
  \, d \varphi ^2) \, - \, dt^2 \;,
\end{equation}
where $a$ is a positive constant, is an example for a traversable wormhole of the 
Morris-Thorne class, see Box 2 in Morris and Thorne \cite{morris-thorne-88}. It 
was investigated, already in the 1970s, by Ellis \cite{ellis-73} who called it a 
``drainhole''. Lensing in the Ellis spacetime was discussed, in a scattering formalism 
assuming that observer and light source are at infinity, by Chetouani and Cl{\'e}ment 
\cite{chetouani-clement-84}. In the following we give a detailed account of lensing 
in this spacetime with the help of our lens equation. 

By comparison of (\ref{eq:wormg}) with (\ref{eq:g}) we find
\begin{equation}\label{eq:wormR}
   A(r) \, = \, 1 \; , \qquad S(r) \, = \, 1 \; , \qquad R(r) \, = \, \sqrt{r^2+a^2} \,  .
\end{equation}
where the radius coordinate $r$ ranges from $r_{\mathrm{min}} = - \infty$
to $r_{\mathrm{max}} = \infty$. (We do \emph{not} identify the region where $r$ is 
positive with the region where $r$ is negative.) The function $R$ has a minimum at 
$r=0$, thereby indicating the existence of circular light rays at the neck of the 
wormhole. In the following we consider the case that observer 
and light sources are on different sides of the neck of the wormhole, 
$- \infty < r_O < 0 < r_S < \infty$. 

\begin{figure}
    \psfrag{aaa}{\small $- \delta _I$} %
    \psfrag{bb}{\small $ \delta _I$} %
    \psfrag{AAA}{\small $-5 \pi$} %
    \psfrag{BBB}{\small $-4 \pi$} %
    \psfrag{CCC}{\small $-3 \pi$} %
    \psfrag{DDD}{\small $-2 \pi$} %
    \psfrag{EE}{\small $- \pi$} %
    \psfrag{F}{\small $\pi$} %
    \psfrag{GG}{\small $2 \pi$} %
    \psfrag{HH}{\small $3 \pi$} %
    \psfrag{II}{\small $4 \pi$} %
    \psfrag{JJ}{\small $5 \pi$} %
    \psfrag{X}{\small $\Theta$} %
    \psfrag{Y}{\small $\Phi$} %
 \epsfig{file=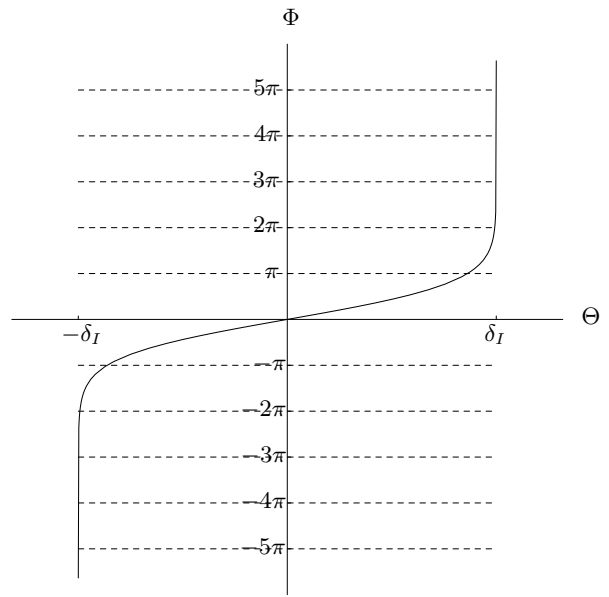,width=0.9\linewidth} 
 \caption{The lens equation (\ref{eq:wormPhi}) for the Ellis wormhole. The
 angle $\delta _I (< \pi /2 )$ is defined in (\ref{eq:wormdelta}). There are 
 infinitely many Einstein rings, indicated by intersections with the dashed 
 lines. \label{fig:worm1}}
\end{figure}

\begin{figure}
    \psfrag{aaa}{\small $- \delta _I$} %
    \psfrag{bb}{\small $ \delta _I$} %
    \psfrag{AAA}{\small $T_0$} %
    \psfrag{X}{\small $\Theta$} %
    \psfrag{Y}{\small $T$} %
 \epsfig{file=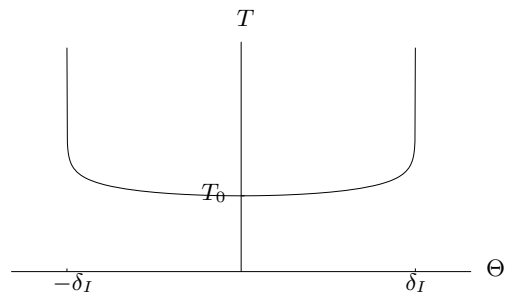,width=0.75\linewidth} 
 \caption{The travel time $T$ as a function of $\Theta$ for the Ellis wormhole. 
  The axis $\Theta =0$ is met at $T_0=r_S-r_O$,
 see (\ref{eq:wormT}).
  \label{fig:worm2}}
\end{figure}

As a first step, we determine for which angles $\Theta$ the lens equation admits
a solution, recall Figure \ref{fig:sky}. In the case at hand, the angles $\delta _I$
and $\delta _{II}$ defined by (\ref{eq:CI}) and (\ref{eq:CII}) are given by
$\mathrm{sin} \, \delta _I = a / \sqrt{r_O^2+a^2}$ and $\delta _{II} = \pi /2 > \delta _I$.
Hence, the lens equation admits a solution for all angles $\Theta$ with
$| \Theta | < \pi /2$ and
\begin{equation}\label{eq:wormdelta}
  {\mathrm{sin}} ^2 \Theta \, < \, \mathrm{sin} ^2 \delta _I
  \, = \, \frac{a^2}{r_O^2+a^2} \; .
\end{equation}
Light sources distributed at $r=r_S$ illuminate a disk of angular radius $\delta _I < 
\pi /2$ on the observer's sky. The apparent rim of the disk corresponds to light rays 
that spiral asymptotically towards $r=0$. As the 
constant $C$ defined by (\ref{eq:C}) satisfies $C \, = \, \sqrt{(r_S^2+a^2)/(r_O^2+a^2)}
\, > \, a/\sqrt{r_O^2+a^2} \, = \, \mathrm{sin} \, \delta _I$, there are no hidden images, 
i.e., the lens equation can be solved for $\Phi$.

\begin{figure}
   \psfrag{AAA}{\small $D_0$} %
   \psfrag{Z}{\small $x$} %
   \psfrag{Y}{\small $D_{\mathrm{ang}}^r , D_{\mathrm{ang}}^t$} %
  \epsfig{file=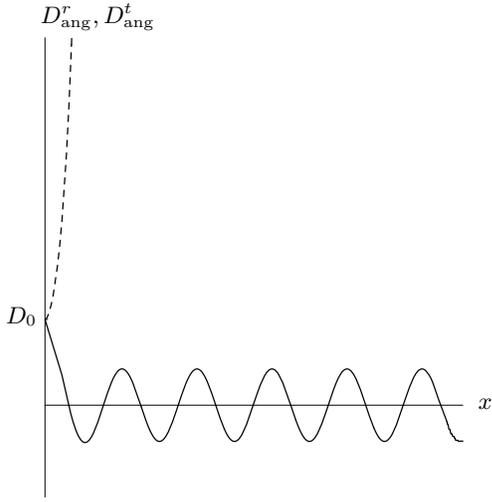,width=0.75\linewidth} 
  \caption{$D _{\mathrm{ang}} ^r $ (dashed) and $D _{\mathrm{ang}} ^t $ 
   (solid) as functions of $\Theta$ for the Ellis wormhole. At each Einstein 
   ring $D _{\mathrm{ang}} ^t $ has a zero. To make the oscillatory behaviour 
   of $D _{\mathrm{ang}} ^t $ visible, we use $x=-\mathrm{log}
   \big(1- | \Theta | / \delta _I \big)$ instead of $\Theta$ on the abscissa; 
   $| \Theta |$ ranges from 0 to $\delta _I$ if $x$ ranges from 0 to $\infty$. 
   The axis is met at  $D_0 = \sqrt{r_S^2+a^2} \sqrt{r_O^2+a^2} \, a^{-1}
   \big( \mathrm{arctan} (r_S /a)- \mathrm{arctan} (r_O /a)\big)$,
   see (\ref{eq:wormDr}) and (\ref{eq:wormDt}). 
   \label{fig:worm3}}
\end{figure}

\begin{figure}
    \psfrag{AA}{\small $0$} %
    \psfrag{BB}{\small $10$} %
    \psfrag{CC}{\small $20$} %
    \psfrag{DD}{\small $30$} %
    \psfrag{Z}{\small $x$} %
    \psfrag{Y}{\small $m=2.5 \, \mathrm{log} _{10} ( D _{\mathrm{lum}} ^2 )+ m_0$} %
   \epsfig{file=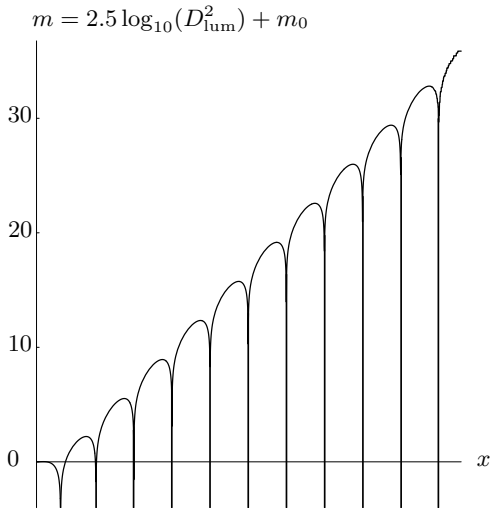,width=0.75\linewidth} 
   \caption{The apparent brightness, measured in terms of magnitude $m= 2.5 \, \mathrm{log} 
   _{10} ( D _{\mathrm{lum}} ^2 )+m_0$, as a function of $\Theta$ for the Ellis wormhole.
   The constant $m_0$ has been chosen such that $m=0$ at $\Theta =0$. 
   We use the same coordinate $x$ as in Figure \ref{fig:worm3} on the 
   abscissa.  At each Einstein ring there is a singularity, $m= - \infty$.
   \label{fig:worm4}}
\end{figure}

With $\Theta$ restricted by (\ref{eq:wormdelta}) we read from (\ref{eq:rs}) that
$\dot{r}$ has no zeros along a ray that starts at $r = r_O$ and passes 
through $r=r_S$. Hence, (\ref{eq:intrs}) gives us $\mathsf{r} (\Theta , s)$ 
and thereby the travel time in terms of an elliptic integral,
\begin{equation}\label{eq:wormintrs}
  s \, = \,  \int_{r_O}^{ \mathsf{r} (\Theta,s)} 
  \frac{\sqrt{r^2+a^2 \,} \, dr}{
  \sqrt{\, r^2 \, + \, a^2 \, \mathrm{cos} ^2 \Theta \, - \, 
  r_O^2 \, \mathrm{sin} ^2 \Theta \, }} \: .
\end{equation}  
Similarly, (\ref{eq:Phi}) gives us $\Phi$ as a (single-valued) function of
$\Theta$ in terms of an elliptic integral and thereby the lens equation,
\begin{equation}\label{eq:wormPhi}
  \Phi \, = \,  \int _{r_O} ^{r_S}  
  \frac{\sqrt{r_O^2+a^2\, } \; {\mathrm{sin}} \, \Theta \; dr}{\, \sqrt{r^2+a^2 \,}  \,
  \sqrt{\, r^2\, + \, a^2 \mathrm{cos} ^2 \Theta 
  \, - \, r_O^2 \, \mathrm{sin} ^2 \Theta \,} \,}  \; ,
\end{equation}  
see Figure \ref{fig:worm1}. As $\Phi$ increases monotonously from $-\infty$ to $\infty$ 
on the domain $\:]- \delta _I , \, \delta _I\, [\,$, there are infinitely many Einstein rings 
whose angular radii converge to $\delta _I$. If we fix a light source at $\varphi = \varphi _0$ 
with $0 < | \varphi _0 | < \pi$, Figure \ref{fig:worm1} gives us infinitely many images which 
can be characterized in the following way. For every $n \in \mathbb{Z}$ there is a 
unique $\Theta _n \in \: \; ] - \delta _I , \, \delta _I \, [ \:$ such that 
$\mathcal{F} (\Theta _n , \varphi _0 + 2n \pi )=0$, and $\Theta _n \to \pm \, \delta _I$ 
for $n \to \pm \, \infty$. 

\begin{figure}
    \psfrag{E}{\small 1} %
    \psfrag{X}{\small $x$} %
    \psfrag{Y}{\small $\varepsilon$} %
 \epsfig{file=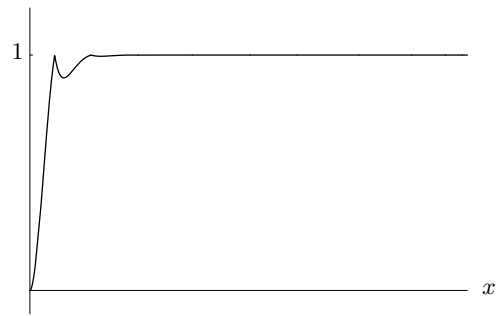,width=0.75\linewidth} 
   \caption{The ellipticity $\varepsilon$ as a function of $\Theta$ for the 
   Ellis wormhole. Again we have chosen the coordinate $x$ from Figure 
   \ref{fig:worm3} on the abscissa. $\varepsilon$ is equal to 1 at the
   Einstein rings and smaller than 1 between the Einstein rings. However, 
   beyond the third Einstein ring $\varepsilon$ stays so close to 1 that 
   in the picture it looks like a constant function.\label{fig:worm5}}
\end{figure}

As $\Phi$ is a (single-valued) function of $\Theta$, so are all observables. 
By evaluating the formulas derived in Section \ref{sec:obs} for the case at 
hand we find
\begin{equation}\label{eq:wormz}
  z \, = \, 0 \: ,
\end{equation}
\begin{equation}\label{eq:wormT}
  T \, = \,  
  \int _{r_O} ^{r_S} \, 
  \frac{\sqrt{r^2+a^2 \,} \; dr}{\, 
  \sqrt{\, r^2\, + \, a^2 \, \mathrm{cos} ^2 \Theta 
  \, - \, r_O^2 \, \mathrm{sin} ^2 \Theta \,} \,} \; .
\end{equation}
\begin{widetext}
\begin{equation}\label{eq:wormDr}
  D_{\mathrm{ang}} ^r \, = \, \int _{r_O} ^{r_S} \,
  \frac{\, \sqrt{\, r_S^2+a^2 \mathrm{cos} ^2 \Theta - r_O^2 \mathrm{sin} ^2 \Theta \, }
  \; \sqrt{\, r_O^2+a^2 \,} \; \sqrt{\, r^2+a^2 \,} \; \mathrm{cos} \, \Theta \; dr}{
  \sqrt{r^2 \, + \, a^2 \, \mathrm{cos} ^2 \Theta \, 
  - \, r_O^2 \, \mathrm{sin} ^2 \Theta \,}^{\; 3}} \: ,
\end{equation}
\begin{equation}\label{eq:wormDt}
  D_{\mathrm{ang}} ^t \, = \, \frac{\sqrt{\, r_S^2+a^2 \,}}{\, \mathrm{sin} \, \Theta \,}
  \: \mathrm{sin} \Big( 
  \int _{r_O} ^{r_S} \, 
  \frac{\sqrt{\, r_O^2+a^2\, } \; {\mathrm{sin}} \, \Theta \; dr}{\, \sqrt{ \, r^2+a^2 \,}  
  \; \sqrt{\, r^2 \, + \, a^2 \, \mathrm{cos} ^2 \Theta 
  \, - \, r_O^2 \, \mathrm{sin} ^2 \Theta \,} \;}
   \Big) \: ,
\end{equation}
\end{widetext}
which gives $D _{\mathrm{lum}}$ and $\varepsilon$ as functions of $\Theta$
via (\ref{eq:Dlum}) and (\ref{eq:epsilon}). The observables are plotted in
Figures \ref{fig:worm2}, \ref{fig:worm3}, \ref{fig:worm4} and \ref{fig:worm5}.

As there are infinitely many Einstein rings whose angular radii converge to 
$\delta _I$, the tangential angular diameter distance must have infinitely
many zeros that converge to $\delta _I$. This is difficult to show in a
picture unless one transforms $\Theta$ into a new coordinate $x$
that goes to infinity for $\Theta \to \delta _I$. If one chooses 
a logarithmic transformation formula, as has been done in Figure \ref{fig:worm3},
one sees that in terms of the new coordinate $x$ the Einstein rings become 
equidistant. This feature is not particular to the Ellis wormhole; 
$\Phi $ as a function of $\Theta$ always diverges logarithmically when a 
circular light ray at a radius $r_p$ with $R'(r_p) = 0$ and $R''(r_p)>0$ is 
approached. The proof can be taken from Bozza \cite{bozza-2002}.

One may also treat the case that observer and light sources are on the same side of 
the neck of the wormhole. If the observer is closer to the neck than the light sources,
$0 < r_O < r_S < \infty$ or $-\infty < r_S < r_O < 0$, the results are quite similar to 
the case above. The only difference is in the fact that the light sources appear as a 
disk of radius \emph{bigger} than $\pi /2$, i.e., the disk covers more than one 
hemisphere. If the observer is farther from the neck than the light sources,
$0 < r_S < r_O < \infty$ or $- \infty < r_O < r_S <0$, there are hidden images. i.e.,
one does not get a single-valued lens map $\Theta \mapsto \Phi$.

The qualitative features of lensing by an Ellis wormhole are very similar
to the qualitative features of lensing by a Schwarzschild black hole. The
radii $r_{\mathrm{min}}=-\infty$, $r_p=0$, $r_{\mathrm{max}}=\infty$ in the Ellis 
case correspond respectively to the radii $r_{\mathrm{min}}=2m$, $r_p=3m$, 
$r_{\mathrm{max}}=\infty$ in the Schwarzschild case. As a matter of fact, 
we encounter these same features whenever the function $R$ has one minimum,
$R'(r_p)=0$ and $R''(r_p) > 0$, and no other extrema on the considered interval.

\section{Concluding remarks\label{sec:concl}}
The lens equation and the formulas for redshift, travel time and radial angular
diameter distance used in this paper refer to lightlike geodesics of the 
$(2+1)$-dimensional metric $A(r)^2 \, \big( \, S(r)^2 \, dr^2 \, + \, 
R(r)^2 \, d \varphi ^2 \, - \, dt^2 \, \big)$, independently of whether this 
metric results from restricting a spherically symmetric and static spacetime 
to the equatorial plane. Therefore, these results apply equally well to 
the plane $z = {\mathrm{const.}}$ of a cylindrically symmetric and static spacetime and, 
of course, to genuinely $(2+1)$-dimensional spacetimes with the assumed symmetries 
such as the BTZ black hole. (For lightlike -- and timelike -- geodesics in the
metric of the BTZ black hole see Cruz, Mart{\'\i}nez and Pe{\~n}a
\cite{cruz-martinez-pena-94}.) E.g., the metric $dr^2 \, + \, k^2 \, r^2 \, d \varphi ^2 
\, - \, dt^2$ results not only by restricting the spacetime of a Barriola-Vilenkin 
monopole to the plane $\vartheta = \pi /2$, as discussed in Subsection 
\ref{subsec:mono}, but also by restricting the cylindrically symmetric and 
static metric $dr^2 \, + \, k^2 \, r^2 \, d \varphi ^2 \, + dz^2 \, - \, dt^2$ 
to the plane $z=0$. The latter metric is well-known to describe the spacetime around 
a static string, see Vilenkin \cite{vilenkin-84}, Gott \cite{gott-85} and 
Hiscock \cite{hiscock-85}, and was investigated in detail already by Marder 
\cite{marder-59,marder-62}. Hence, if re-interpreted appropriately, the 
results of Subsection \ref{subsec:mono} apply to light rays in the plane 
perpendicular to a static string. For treating \emph{all} light rays in
a cylindrically symmetric and static spacetime one may introduce a
modified lens equation, replacing the sphere $r=r_S$ with a cylinder.


\end{document}